\documentclass[eqsecnum,aps]{revtex4}

\begin{document}

\title{Comment on Repulsive Casimir Forces}

\author{Davide Iannuzzi and Federico Capasso\footnote{e-mail:
capasso@deas.harvard.edu}}
\address{Harvard University, Division of Engineering and Applied Sciences, Cambridge, MA 02138}
\author{PACS numbers: 12.20.Ds, 85.85.+j}

\maketitle

A recent theoretical calculation shows that the Casimir force between two
parallel plates can be repulsive for plates with nontrivial magnetic
properties\cite{prl}. According to the authors, the effect may be observed
with known materials, such as ferrites and garnets, and it might be
possible to engineer micro- or nanoelectromechanical systems (MEMS or
NEMS) that could take advantage of a short range repulsive force. Here we
show that on the contrary the Casimir force between two parallel plates in
vacuum at micron and submicron distance is always attractive. 

The Casimir energy $E$ per unit area of two infinite slabs of materials with
permittivity and permeability $\epsilon_{1,2}$, $\mu_{1,2}$ maintained at
a distance $a$ in vacuum is given by equation (4) of \cite{prl}. To obtain
equation (4), the authors parametrize the field modes using the vector
{\bf k} $=(k_x,k_y,k_t)$. In their notation, $k_x$ and $k_y$ are the
components of the wave vector parallel to the plates, and $k_t$ is defined
by the Wick rotation $k_t \leftrightarrow i \omega$, where $\omega$ is the
frequency. Integrating equation (4) over $\vert${\bf k}$\vert$ to obtain a simpler equation for $E$, the authors implicitly that $\epsilon_{1,2}$ and $\mu_{1,2}$ do not
depend on $\omega$.  From equation (7), the authors derive the following
conclusions: 

\noindent (i) When both materials have high $\mu$ and $\epsilon$,
the sign of the force depends on the impedances of the materials. 

\noindent (ii) When one of the two bodies is a perfect conductor, and the
other has large $\mu$ and $\epsilon$, the sign of the force depends on the
impedance of the latter. 

\noindent (iii) In the uniform velocity of light (UVL) case, the sign of
the force depends on the values of $\mu$ of the two materials.

Finally the authors claim that a class of materials with high
permeability might be suitable for a demonstration of the repulsive
Casimir force, which could play a crucial role in the construction of MEMS
and NEMS.

All the conclusions depend on the authors' physically
incorrect assumption that $\epsilon_{1,2}$ and $\mu_{1,2}$ are independent of
$k_t$. 

While the assumption that $\epsilon$ and $\mu$ are independent of $\omega$
is incorrect, one can examine the question whether in the range of
frequencies which give the largest contribution to the Casimir force
(corresponding to wavelengths $\sim a$) this is indeed a reasonable
approximation or not. The plate-sphere\cite{ps}, cylinder-cylinder\cite{cy} or plate-plate\cite{pp}
separations used in Casimir force measurements, as well as the typical
distances between mechanical parts in MEMS and NEMS and those relevant to
stiction problems, are in the submicron to micron range, corresponding to
frequencies in the optical to mid-infrared region. In this spectral range
definitely $\epsilon$ cannot be assumed to be weakly dependent on
frequency for most materials. Furthermore, concerning the hypothesis on
$\mu_{1,2}$, it can be shown from theoretical arguments that in the
optical range the permeability of materials ceases to have physical
meaning\cite{landau}. In practice, no known material presents significant
deviations from $\mu=1$ in a much wider spectral range. As a consequence,
for separations in the micron and submicron range, equation (4) becomes
the usual Lifshitz formula\cite{lifshitz}, from which it follows that the Casimir force between is
always attractive.

Interestingly another recent article\cite{prlbruno} shows that for two
ferromagnetic slabs the magneto-optical Kerr effect gives rise to an
additional attractive {\it Casimir magnetic force}. The result is obtained
using an equation similar to equation (4) of \cite{prl}. In this case,
however, it is implicitly assumed that $\mu=1$, and the force is
calculated by only considering the dependence of the dielectric tensor on
$\omega$.

In conclusion we believe that the calculations presented in \cite{prl} are
not suitable for drawing conclusions on the sign and the magnitude of the
Casimir force between real device elements at distances relevant to Casimir force
measurements and to nanomachinery.  At these distances the
Casimir force between two slabs in vacuum is always attractive.

We thank C. Varma for useful discussions.

\end{document}